# TOWARDS INTEGRATED CLINICAL-COMPUTATIONAL NUCLEAR MEDICINE

Faraz Farhadi, MD[1,2]; Shadi A. Esfahani, MD, MPH[1]; Fereshteh Yousfirizi, PhD[3]; Monica Luo, BSc[3]; Pedro Esquinas Fernandez, PhD[4]; Arkadiusz Sitek, PhD[1]; Hamid Sabet, PhD[1]; Babak Saboury, MD, MPH, DABR, DABNM[2, 3, 5]; Arman Rahmim, PhD, DABSNM[3 ,6 ,7]; Pedram Heidari, MD[1*]

1 Department of Radiology, Massachusetts General Hospital, Boston, MA 02114, USA
2 Department of Nuclear Oncology, Institute of Nuclear Medicine, Bethesda, MD 20815, USA
3 Department of Integrative Oncology, BC Cancer Research Institute, 675 West 10th Avenue, Vancouver, BC V5Z 1L3, Canada
4 Molecular Imaging and Therapy, BC Cancer, Vancouver, BC, Canada.
5 Department of Radiology and Imaging Sciences, Clinical Center, National Institutes of Health (NIH), 9000 Rockville Pike, Bethesda, MD 20892, USA
6 Department of Physics & Astronomy, University of British Columbia, Vancouver, BC, Canada.
7 Department of Radiology, University of British Columbia, 675 West 10th Avenue, Vancouver, British Columbia V5Z 1L3, Canada

**\*Corresponding author**
Pedram Heidari, MD
Nuclear Medicine and Molecular Imaging
Department of Radiology, Massachusetts General Hospital
55 Fruit St
White 427
Boston, MA 02114, USA

**Abstract**

The field of Clinical-Computational Nuclear Medicine is rapidly advancing, fueled by AI, tracer kinetic modeling, radiomics, and integrated informatics. These technologies improve imaging quality, automate lesion detection, and enable personalized radiopharmaceutical therapy through physiologically based pharmacokinetic (PBPK) modeling and voxel-level dosimetry. Workflow automation and Natural Language Processing (NLP) further enhance operational efficiency. However, successful implementation and adoption of these tools require clinical oversight to ensure accuracy, interpretability, and patient safety. This paper highlights key computational innovations and emphasizes the critical role of clinician-guided evaluation in shaping the future of precision imaging and therapy.

**Key Points:**

1. Clinical-Computational Nuclear Medicine integrates AI, radiomics, and kinetic modeling to improve imaging, automate lesion detection, and personalize treatments.
2. AI tools like Natural Language Processing (NLP) are advancing automated reporting, but clinical use is still experimental.
3. Patient-specific dosimetry and PBPK modeling support precision therapy by optimizing doses and reducing toxicity.

4. Workflow automation, Health Level Seven International (HL7) and Fast Healthcare Interoperability Resources (FHIR), and analytics improve efficiency and reduce errors.
5. The future relies on clinical oversight, validation, and human-in-the-loop approaches for safety, reliability, and accuracy.

**Introduction**

Nuclear medicine is inherently computational, setting it apart from many other medical specialties since its inception.[1] This computational foundation stems from nuclear medicine's core physics, which relies on quantitative frameworks to analyze physiological and molecular signals.[5] The term 'Computational Nuclear Medicine' includes various algorithmic, modeling, and informatics techniques used to support clinical practice, such as artificial intelligence (AI), Monte Carlo simulations, tomographic image reconstruction, tracer kinetic modeling, These approaches enhance accuracy, objectivity, and scalability, enabling clinicians to manage the growing data volume and complexity in modern diagnostics and therapy.[2–4]

Modern nuclear medicine operates within a computational environment driven by hybrid imaging, high-dimensional data, and by expanding theranostic applications. Radiopharmaceutical workflows go beyond diagnostics, utilizing computational tools for treatment planning, dosimetry, and outcome prediction. Radiomics and AI further improve these techniques, requiring scalable, robust infrastructure for clinical implementation. The FDA has approved over 1,240 AI-enabled medical devices, including more than 950 radiology-related tools (Reviewed on August, 2025).[22]

Computational tools in nuclear medicine support clinical decision-making and improve workflow efficiency. Decision-support technologies influence management by analyzing imaging and clinical data.[6] Examples include AI-based reconstruction and denoising to improve image quality, and enable dose reduction;[3] automated lesion detection, segmentation, and classification tools for objective disease assessment,[7] and patient-specific dosimetry using pharmacokinetic modeling to optimize Radiopharmaceutical Therapy (RPT) planning.[8] Workflow optimization tools enhance care delivery, resource use, and interoperability without directly affecting clinical decisions. Examples include automated Appropriate Use Criteria (AUC) compliance and patient scheduling systems,[9] workflow automation platforms employing Health Level Seven International (HL7) and Fast Healthcare Interoperability Resources (FHIR) standards for streamlined scheduling and radiotracer logistics;[10] Natural Language Processing (NLP) tools for electronic medical record mining, automated report generation, and question-answering tasks;[11] and predictive analytics models aiding operational planning, such as forecasting scanner use and managing radiation safety.[9]

Integrating these computational advances into clinical practice requires rigorous validation. The SNMMI AI Task Force's RELAINCE (Reliability, Experimentation, Lack of bias, Interpretability, Authenticity, Non-inferiority, Communication, and Evaluation) framework assesses AI tools by scientific potential, technical performance, clinical impact, and post-deployment monitoring.[43] Regulatory progress continues with FDA approval of several platforms.[44] Comprehensive validation includes retrospective studies, prospective trials, and post-deployment monitoring to confirm real-world performance. Addressing discrepancies between algorithm outputs and clinical expectations is critical. Robust evaluation must span diverse patients, imaging protocols, and settings to ensure applicability and trust. While these validation guidelines are important, transparent reporting of model development, training data, validation methods, and performance metrics is essential for clinical acceptance.[43]

While advancing nuclear medicine through improved accuracy, consistency, and error reduction is important, the central goal is to ensure physicians remain in the loop so that computational tools are applied in ways that ultimately improve patient care. Computational systems can process vast data and detect subtle patterns beyond human capability, offering improved accuracy and consistency. However, these models may still encounter limitations such as bias, unexpected errors, or context misinterpretation that require expert clinical judgment to identify and correct. Keeping physicians engaged ensures that AI outputs are interpreted within the broader clinical context, preserving patient safety and ethical responsibility. This balance uses the strengths of both AI and human expertise, fostering trust and ensuring that technology acts as a reliable aid, not a replacement in patient care.

Here, we highlight key areas at the forefront of clinical-computational nuclear medicine, along with emerging fields that show future potential, as illustrated in Figure 1. This is not an exhaustive list. In addition, we discuss the importance of physician oversight and integration in these fields to ensure the validity and effectiveness of these technologies in real-world scenarios.

## Computational Tools in Theranostics

Computational tools in theranostics integrate advanced data analysis and imaging techniques to guide both diagnosis and therapy, enabling precise, patient-specific treatment by combining diagnostic insights with therapeutic planning. In nuclear medicine, especially regarding AI-driven outcome prediction, there remains a need for prospective studies or randomized controlled trials (RCTs), systematic reviews, and meta-analyses to validate theranostics technologies in clinical practice[12]. To foster adoption, high-quality evidence of clinical efficacy is essential. The following sections discuss key technological domains, including image reconstruction and denoising, lesion detection and quantitative PET biomarkers, radiomics-based outcome prediction, tracer kinetic modeling, patient-specific dosimetry, and protocoling for advancing theranostics.

A. Image Reconstruction and Denoising

Advances in nuclear imaging have improved image reconstruction and post-processing to address noisy, low-count PET and SPECT with limited resolution from physical constraints. Traditional methods like Ordered-Subset Expectation-Maximization (OSEM) optimize scan time or tracer activity. More recently, AI methods, particularly Deep Learning (DL), have greatly enhanced image quality through denoising, super-resolution, and artifact correction techniques,[3] supporting streamlined clinical decisions. For instance, Convolutional Neural Networks (CNNs), U-Net architectures, Generative Adversarial Networks (GANs), and Vision Transformers (ViTs) have significantly improved low-count or low-dose imaging quality. These advancements enable shorter PET/SPECT scans or lower doses while maintaining diagnostic accuracy.[13–15] Reduced injected activity decreases radiation exposure, and shorter scan times improve patient comfort, reduce motion artifacts, and increase patient throughput. For example, DL techniques successfully denoised half-count PET scans, achieving lesion detectability comparable to standard full-dose scans.[13,16]

Unsupervised and physics-informed AI methods are emerging, addressing supervised methods' reliance on matched image pairs. Physics-Informed Neural Networks (PINNs)[17] integrate physical principles, such as decay kinetics and photon transport, into their architectures. This enhances generalizability and accuracy while minimizing the need for extensive labeled datasets,[18] promising increased robustness in clinical settings.

Achieving high image fidelity requires addressing technical challenges, such as resolution recovery, attenuation and scatter correction, and motion correction.[20] DL-generated attenuation correction maps from emission data or anatomical priors can potentially eliminate additional CT scans, reducing radiation exposure and streamlining workflows.[21] In dynamic cardiac PET, DL-based bidirectional Long Short-Term Memory (LSTM) networks effectively correct inter-frame motion, surpassing traditional techniques by considering tracer kinetics.[20]

Despite these advantages, AI denoising raises concerns regarding diagnostic accuracy. For example, clinical evaluation of AI-denoised low-dose $^{64}$Cu-DOTATATE PET scans showed high image quality but compromised lesion detection, notably missing lesions visible in full-dose scans, as illustrated in Figure 2.[19] Advanced unsupervised and PINNs can preserve diagnostic confidence and, with continued validation, are poised for integration into clinical workflows for real-time image enhancement.

B. Lesion Detection, Segmentation, and Conventional Quantitative Biomarkers

Lesion detection and segmentation for tumor and organ uptake, traditionally done manually, is labor-intensive and prone to observer variability. Quantitative biomarkers like Metabolic Tumor Volume (MTV) and Total Lesion Glycolysis (TLG) offer objective, reproducible prognostic value. MTV thresholds have informed treatment strategies in endometrial cancer and lymphoma.[39,40] Additionally, combining Total Metabolic Tumor Volume (TMTV) with clinical factors has outperformed traditional risk models.[41] Although TMTV is a critical measure of tumor burden and therapeutic response, it is currently limited by manual processes and variability.[28,29] Despite its importance, inconsistent segmentation practices hinder adoption. Recent AI advances now automate detection, segmentation, and biomarker extraction, enabling consistent, reliable clinical assessments.[23] For example, DL architectures such as 3D U-Nets, GANs, and ViTs have demonstrated expert-level performance in identifying hypermetabolic lesions in $^{18}$F-fluorodeoxyglucose (FDG)-PET across various cancers, generating clinically relevant metrics including MTV, TLG, and standardized uptake values (SUV).[27] In breast cancer, DL-based tools achieved sensitivity and specificity nearing expert levels.[23] These DL models achieve 80% compliance with the standardized TMTV_benchmark protocol, ensuring uniformity without manual intervention and outperforming conventional methods by reducing false positives. (Figure 3). As with any AI algorithm, these performance metrics should be interpreted in the context of external validation, as sources of error remain. For example, reported accuracy can vary substantially depending on whether ground truth segmentations are clinician-defined from scratch or generated by AI and subsequently edited.

**C. Next-Generation Quantification: Radiomics for Outcome Prediction**

Expanding on quantitative imaging principles, AI-integrated radiomics enables extraction and interpretation of high-dimensional biomarkers for predictive modeling. While MTV and TLG focus on metabolic volume and uptake, radiomics captures tumor shape, texture, intensity, and heterogeneity from PET and SPECT images. When combined with Machine Learning (ML) and DL, these radiomic features power predictive models for diagnosis, treatment response assessment, and outcome prediction. AI methods automate biomarker extraction, demonstrating accuracy across cancers using various tracers.[30–36] Despite its promise, radiomics lacks biological interpretability and clinical validation through prospective trials. Most studies remain retrospective, limiting their adoption in evidence-based medicine.[58] Nonetheless, DL-radiomics holds potential for advancing clinical decision support in therapy response prediction, toxicity assessment, trial stratification, and personalized dosimetry. By capturing tumor biology beyond SUVmax, DL-driven approaches outperform conventional metrics. In non-small cell lung cancer (NSCLC), CNNs trained on PET radiomics outperform conventional metrics in predicting therapy response and survival.[49] Beyond oncology, radiomics aids in cardiology, predicting coronary artery disease severity via SPECT/PET[51] and neurology, where FDG-PET-based models better distinguish early Alzheimer's from mild cognitive impairment [52] and predict cognitive decline using metabolic brain PET patterns.[42] These models detect non-linear patterns[49] and leverage DL architectures to extract high-level imaging biomarkers,[49,53] surpassing traditional ML methods.[54]

Predictive modeling in nuclear medicine aims to forecast outcomes to guide therapy. It begins with defining the clinical question and selecting endpoints (e.g., survival, quality of life, or response rate), each shaping interpretability and statistical handling, especially with censoring.[46] Model choice depends on the endpoint; censored time-to-event data often use survival analysis like Cox regression, ML models like random survival forests, or DL models such as DeepSurv, while binary outcomes may rely on neural networks, ML models like XGBoost, or transformers.[47] Feature selection and evaluation must align with study goals, combining clinical insight with methods like LASSO. Cross-validation and SHAP (Shapley Additive exPlanations) values with randomized data help ensure robustness and prevent spurious associations.[48]

While predictive modeling frameworks are advancing, the choice of radiomic inputs is equally critical. Conventional measures such as SUVmax and hand-crafted (shallow) radiomics defined by Image Biomarker Standardisation Initiative (IBSI) have been extensively studied but show limited generalizability. Deep radiomics, derived from CNNs or transformers, capture complex, non–human-engineered features and often outperform shallow approaches in segmentation and prediction. However, these methods require large, harmonized datasets that remain scarce in nuclear medicine. Consequently, shallow radiomics remain in use, though few are clinically validated or approved.

Multimodal integration of imaging, clinical, and genomic data further enhances prediction and personalization.[49] Yet, clinical translation faces barriers like poor feature reproducibility, data heterogeneity, and the 'black-box' nature of AI models.[55] Addressing these requires protocol standardization, harmonized pipelines, external validation,[56] and the adoption of explainable AI to build clinical trust.[49] Guidelines like IBSI[57] support standardization, while federated learning addresses privacy and scalability concerns.

Despite these advances, important caveats remain. Conventional metrics like SUVmax can be viewed as shallow radiomic features under IBSI, alongside other hand-crafted descriptors that have shown limited generalizability over the past 15 years and have yet to yield FDA-approved, clinically adopted tools. Deep radiomics using CNNs or ViTs have the potential to offer predictive power but requires large, curated datasets that nuclear medicine rarely possesses, which restricts its impact beyond tasks such as segmentation.

**D. Tracer Kinetic Modeling and Parametric Imaging**

Tracer kinetic modeling quantifies physiology from dynamic radiopharmaceutical distribution, while parametric imaging maps voxel-wise parameters such as metabolic, receptor binding, and perfusion. Despite their clinical utility, adoption is limited by complexity, acquisition time, and computational demands.

DL addresses these limitations by linking time-activity curves and physiological parameters, reducing reliance compartmental models and input functions. Supervised learning estimates kinetic parameters such as the influx constant (Ki) or volume of distribution (Vd) directly from dynamic PET data, enabling faster and more robust parametric imaging.[59] Recent methods using 3D U-Net architectures and self-supervised learning further improve voxel-wise parameter estimation by integrating physiological constraints into training objectives.[60]

An innovative application of kinetic model-informed DL is in multiplexed PET image separation. This methodology leverages established kinetic behavior to disentangle signals originating from multiple radiotracers administered either concurrently or sequentially, otherwise indistinguishable, employing conventional reconstruction techniques.[61] Such techniques demonstrate potential for dual-tracer protocols, including the combination of perfusion and receptor-targeted agents. The resultant multiplexed parametric maps can offer deeper biological insights and improve workflow efficiency in research and clinical settings.

AI also facilitates the estimation of input functions based on data-driven approaches, addressing a traditional bottleneck in kinetic modeling. Models trained on population or multimodal data generate subject-specific input functions without arterial sampling.[62] Similarly, likelihood-free Bayesian inference utilizing neural networks[63] has facilitated the estimation of Physiologically Based Radiopharmacokinetic (PBRPK) parameters from incomplete time activity curves, thereby supporting the reconstruction of input functions in scenarios characterized by limited data.

Once limited to research parametric imaging, is no important for therapy monitoring and individualized treatment planning. CNNs trained on dynamic PET data can generate Ki parametric maps with accuracy comparable to Patlak analysis, while requiring less computation and improving noise resistance.[64] This significantly lowers barriers to clinical use of dynamic imaging in oncology, cardiology, and neurology. Specifically, in RPT, AI-generated parametric maps facilitate the quantification of radiotracer retention and clearance kinetics at the voxel level, thereby supporting personalized dosimetry and response prediction.

While these AI methods demonstrate significant potential, several challenges persist - specifically, the interpretability of models, their generalizability across various radiotracers and imaging scanners, and the necessity for standardized dynamic datasets to facilitate validation.

It is important to recognize that images are ultimately human constructs, generated under assumptions about compartmental models and input functions. These assumptions can introduce noise amplification and systematic bias, potentially limiting accuracy. AI offers an alternative path by bypassing these intermediates, learning predictive representations directly from dynamic or even non-reconstructed data. Such approaches not only avoid error propagation from parametric image generation but also leverage the raw temporal signal in a way that aligns more closely with how AI models extract information.

### E. Patient-Specific Dosimetry and Treatment Planning Tools

Patient-specific dosimetry is a fundamental element of precision RPT, facilitating the enhancement of therapeutic effectiveness while reducing toxicity to non-target tissues. In contrast to external beam radiation therapy, where dose delivery can be accurately shaped and monitored in real-time, RPT depends on biologically distributed radiation, rendering the estimation of absorbed doses intrinsically more complex. Conventional dosimetry methods have frequently employed population-based models or simplified assumptions regarding radiotracer distribution and kinetics, which often result in suboptimal treatment customization.

Recent advances in computational tools and imaging technologies are revolutionizing this paradigm. Voxel-based dosimetry, supported by quantitative SPECT/CT or PET/CT, facilitates the calculation of three-dimensional dose distributions at the level of individual organs, tumors, and subvolumes. Monte Carlo simulation platforms, such as GATE and 3D-RD, have historically been employed to model particle transport and energy deposition; however, their clinical adoption has been constrained by high computational demands and complex workflows. Currently, there are FDA-approved commercial software solutions, such as Voximetry Torch, that efficiently perform Monte Carlo-based dosimetry estimations.[65] Additionally, AI-enabled approaches are streamlining these processes. For example, ML and DL models can accelerate accurate dose estimation, automate the segmentation of organs and lesions, and predict time-integrated activity maps from limited time-point imaging.[66,67]

Patient-specific dosimetry frameworks increasingly incorporate PBPK modeling to more accurately reflect interpatient variability in radiopharmaceutical distribution and clearance. These models utilize prior knowledge of tracer kinetics, combined with patient-specific imaging data, to simulate time-activity curves and estimate absorbed doses to target tissues and organs at risk. Such tools are particularly relevant in therapies utilizing beta-emitters (e.g., $^{177}$Lu-DOTATATE) and alpha-emitters (e.g., $^{225}$Ac-labeled agents), where dose-response relationships are critical for treatment success and toxicity prediction.[68–70] A PBPK-informed cGAN predicts voxel-wise post-therapy dosimetry from pre-therapy PET by using physiologically derived dose constraints as priors. Instead of directly estimating PBPK parameters, it leverages simulated $^{68}$Ga-PSMA-11 PET, segmentation masks, and organ-specific PBPK dose ranges to generate 3D voxel-level maps for $^{177}$Lu-PSMA-I&T therapy. By merging mechanistic pharmacokinetic modeling with DL, it yields biologically plausible, spatially heterogeneous dose distributions, improving dose-volume histogram accuracy and reducing organ-level errors for personalized RPT planning.[68]

These tools incorporate quantitative imaging, dosimetry calculations, and predictive models of normal tissue complication or tumor control probability to facilitate personalized treatment prescriptions. For instance, voxel-level dose-volume histograms can be generated to inform activity selection and scheduling, while radiomics and ML algorithms may assist in forecasting therapeutic responses based on baseline imaging features.[71]

Automation and standardization are key for integrating these tools into clinical workflows. AI-driven segmentation, image registration, and kinetic modeling reduce interobserver variability and improve reproducibility. Seamless interoperability with Picture Archiving and Communication System (PACS), treatment systems, and regulatory frameworks is essential to embed these solutions in practice. As a result, patient-specific dosimetry is poised to become a standard RPT component, enabling evidence-based, personalized planning.

### F. Protocoling and Imaging Appropriateness

Computational Decision Support Systems (DSS) represent a key advancement in clinical practice. Nuclear medicine physicians often guide referring clinicians on appropriate imaging and therapy based on evidence-based criteria. Computational systems help ensure AUC compliance and optimize patient selection, transforming how imaging appropriateness is assessed and maintained.

The Centers for Medicare & Medicaid Services (CMS) mandate under the Protecting Access to Medicare Act of 2014 (PAMA) requires that advanced diagnostic imaging orders, including PET and other nuclear medicine studies, consult AUC through a Clinical Decision Support Mechanism (CDSM).[72] This requirement has driven the development of software that integrates with electronic ordering systems to give real-time feedback on imaging appropriateness. These platforms use rule-based algorithms or AI trained on guidelines and patient data to recommend suitable tests and flag low-value studies.[73] The

American College of Radiology (ACR) Appropriateness Criteria cover 257 topics and over 4,000 clinical scenarios as the evidence base for these systems.[74]

Rule-based algorithms utilize conditional statement structures that assess patient-specific clinical parameters against established criteria.[75] These deterministic systems offer advantages in terms of generalizability, modularity, and ease of implementation, while preserving explainability essential for clinical acceptance. More sophisticated implementations incorporate ML approaches capable of NLP within electronic health records and matching patient characteristics to eligibility criteria with increasing sophistication.[76] The selection of patients for RPT exemplifies a particularly advanced application of automated decision support, wherein algorithmic systems can simultaneously verify multiple eligibility criteria,[77] including imaging results from PSMA-PET scans, laboratory parameters, treatment history, and contraindications, thereby providing a comprehensive assessment of patients in real-time.[78] As these computational methodologies mature, they are poised to become essential components of nuclear medicine practice, ensuring optimal patient selection while streamlining clinical workflows and alleviating administrative burdens on healthcare providers.[9]

## Computational Tools in Workflow Optimization and System Efficiency

### A. Workflow Automation and Integration (HL7/FHIR and Beyond)

Nuclear medicine departments are complex ecosystems requiring coordination of scheduling, radiopharmaceutical logistics, scanner use, reporting, and regulatory compliance. Integrating workflow automation and clinical informatics is essential to improve efficiency, reduce human error, and meet the growing demands of healthcare.

Computational algorithms increasingly optimize radiopharmaceutical logistics. Real-time scheduling systems using dynamic control methods have reduced PET exam times by an average of 6.1 minutes per patient while minimizing errors by aligning scans with cyclotron production and patient flow.[79] In radioembolization, operations scheduling software aligning catheter lab availability with dose preparation has cut patient wait times by 70% and reduced staff radiation exposure by minimizing idle handling of high-activity materials.[80] These applications highlight the utility of linear programming, heuristic algorithms, and logic-based techniques for automating complex, constraint-driven workflows.

Fully digital workflows also support ongoing performance monitoring and analytics. Timestamping key events like radiotracer administration and scanner use enables frameworks like queuing theory and predictive modeling to identify bottlenecks and optimize radiotracer supply chains.

Effective automation and analytics require robust interoperability standards for structured data exchange across clinical systems. HL7 Version 2 remains the core framework for order communication and results reporting, ensuring legacy compatibility. FHIR introduces modern APIs and standardized resources (e.g., ImagingStudy, DiagnosticReport) for modular integration within enterprise IT. For departments aiming to scale precision workflows and system-level intelligence, adopting these frameworks is essential; without them, automation remains siloed and analytics fragmented, limiting the potential of computational nuclear medicine.

### B. Natural Language Processing and Automated Reporting

Integrating NLP with computer vision holds promise for clinical nuclear medicine but remains mostly at the proof-of-concept stage, with limited feasibility studies. PET and SPECT reports are traditionally dictated with unstructured language and variable styles. While AI-driven report generation and lesion labeling are emerging, these applications remain largely research-based and are not yet ready for routine clinical use.[81–83]

Automated reporting improves consistency in longitudinal assessments, reduces radiologist workload, ensures template compliance, and enables real-time, machine-readable data for decision support and research.

Beyond report generation, NLP plays a central role in retrospective data extraction and informatics. Algorithms can parse large volumes of free-text nuclear medicine reports to identify structured elements such as lesion counts, radiotracer uptake values, and treatment responses. This facilitates population-level analytics, quality benchmarking, and predictive modeling. When integrated with electronic health records (EHRs), NLP tools can contextualize imaging findings by automatically retrieving relevant clinical history, laboratory data, or therapeutics, thereby further enhancing interpretive accuracy and clinical relevance.[82]

Despite these advances, challenges persist. Variability in reporting styles, limited availability of large, annotated datasets, and the necessity for cross-institutional generalization remain persistent barriers. Consequently, current applications are primarily at the proof-of-concept stage, with only a limited number of studies demonstrating feasibility.[84–86] These models are vulnerable to hallucinated findings, biases, and misinterpretation of subtle imaging features, emphasizing the importance of human-in-the-loop oversight. In this framework, nuclear medicine physicians validate AI outputs, rectify errors, and offer expert feedback that iteratively enhances model performance while maintaining diagnostic integrity.[85]

Generative AI should assist - not replace - clinicians, supporting high-level interpretation while learning from their input to improve. Still, validation, bias detection, and strong governance are essential to prevent workflow disruption or cognitive overload.[87]

## C. Resource Optimization and Safety Monitoring

Nuclear Medicine's operational demands, including costly imaging infrastructure, radiation safety, and radiopharmaceutical logistics, require data-driven strategies for resource optimization, safety, and quality. Advances in computational methods, especially ML and predictive analytics, offer scalable solutions by enabling proactive decision-making and reducing inefficiencies.

ML algorithms support predictive scheduling in nuclear medicine by analyzing historical attendance, no-show rates, and resource use to forecast imaging demand. Gradient-boosted regression tree models can predict no-shows with high accuracy, enabling targeted interventions for high-risk patients and reducing missed appointments.[88] For instance, the clinical deployment of the XGBoost-based scheduling system has been associated with a 17.2% reduction in no-shows and a 7% increase in scanner throughput, enhancing the operational efficiency of PET/CT infrastructure.[6] The SNMMI AI Task Force has identified such predictive models as crucial tools for optimizing patient scheduling, resource allocation, and equipment utilization.[9]

These systems facilitate early detection of performance deterioration and help prevent unplanned outages. By integrating data from Internet of Things (IoT) sensors and real-time performance metrics, including temperature fluctuations, image quality indicators, and equipment status parameters, these systems establish operational baselines and identify potential failures.[89] Implementation of predictive maintenance has achieved a 42% reduction in unplanned downtime and extended the lifespan of equipment through proactive scheduling of interventions during off-peak hours.[6] ML models can process data from hundreds of electronic sensors to comprehensively monitor equipment and identify early failure warning signs with high precision.

Radiation safety management has been enhanced through application of computational analytics to dosimetry data and exposure monitoring. AI-driven anomaly detection systems applied to personal dosimeter readings have demonstrated 98% sensitivity in identifying unusual exposure events, allowing for timely activation of intervention protocols.[90] ML models, particularly random forest regression, have shown improved accuracy in dosimeter reading interpretation compared to conventional algorithms, with notable gains in the context of low-energy photon exposures.[91] In parallel, association rule mining has been utilized to analyze large-scale dosimeter datasets, revealing patterns that link specific procedural variations, work practices, and temporal trends to an elevated radiation exposure risk, thereby informing targeted safety interventions.[90,92]

System-level optimization in nuclear medicine has been advanced through the use of digital planning tools and simulation models that support strategic resource management and workflow efficiency. Scheduling frameworks based on Answer Set Programming have demonstrated effective patient allocation by accounting for disease-specific imaging needs, radiopharmaceutical availability, and scanner capacity constraints. Radiopharmaceutical inventory management systems incorporating computerized algorithms automatically apply decay correction, track storage conditions, and maintain full traceability of all transactions, significantly reducing operational errors and minimizing waste.[93] Ongoing developments aim to integrate ML approaches for individualized dosimetry prediction and to refine scheduling algorithms to accommodate increasing patient volumes while preserving safety and operational performance standards.[94]

## Physician-in-the-Loop: Clinical Oversight and Integration

These computational advances are optimal when integrated into a clinician-in-the-loop framework, wherein physicians actively validate and refine AI-generated outputs. Clinicians' interactions (e.g. accepting, rejecting, or modifying AI suggestions) enable continuous model improvement through active learning. This collaboration leverages physicians' clinical expertise, aligning AI outcomes with patient-specific contexts [24] Additionally, this interactive process benefits medical education, though care must be taken to prevent increased physician workload or burnout.[25,26] Integrating clinician oversight ensures computational tools remain clinically and ethically sound, promoting augmented intelligence rather than autonomous operation.[24] AI algorithms risk introducing subtle artifacts or "hallucinations" that could compromise diagnostic accuracy. This underscores the necessity of task-based evaluations to ensure that lesion detection accuracy remains uncompromised.

To facilitate clinical integration, reducing annotation burdens on clinicians is vital. While physician-in-the-loop involvement remains essential for validation and oversight, techniques like transfer, active, and self-supervised learning help minimize labeling demands without compromising accuracy.[37] Active learning selectively requests clinician annotations for uncertain cases, significantly reducing workloads.[38] Semi-supervised transfer learning framework can successfully segment PET/CT data with limited annotations (Figure 4).[27]

Implementation science frameworks like integrated knowledge translation (IKT) promote meaningful clinician engagement from AI development through deployment. Clinicians and stakeholders collaborate with developers to refine problem definitions, metrics, strategies, and outcomes. This integration ensures AI tools address real clinical needs, bridging the gap between innovation and practice.[45]

## Summary and Conclusions

Nuclear medicine is undergoing a major transformation through the integration of advanced computational techniques and clinical practice. This shift, known as Clinical-Computational Nuclear Medicine, is redefining the planning, delivery, and evaluation of nuclear imaging and RPT. Innovations such as AI, kinetic modeling, radiomics, and informatics are driving a move from qualitative, population-based methods to quantitative, patient-specific, adaptive approaches.

Key innovations include AI-based image reconstruction and denoising to improve diagnostic quality while reducing dose and scan time; automated lesion detection and segmentation for reproducible, high-throughput quantification; and radiomics-based models to guide personalized therapy and response assessment. Concurrently, kinetic modeling and AI-driven parametric imaging are reestablishing PET and SPECT as clinical tools for tracer quantification and treatment planning.

In therapy, patient-specific dosimetry and PBPK-based planning tools enable precision RPT by optimizing dose delivery and minimizing adverse effects. These advances are supported by clinical decision support, workflow automation, and NLP-assisted reporting, which ensure appropriateness, improve efficiency, and reduce administrative burden.

The success of these technologies depends on active engagement from nuclear medicine physicians or other related specialists, who bridge algorithmic outputs and clinical decisions. Their expertise ensures AI tools are validated, refined, and integrated into personalized care by augmenting, not replacing, clinical judgment. These tools form a cohesive ecosystem positioning nuclear medicine as a model of data-driven, precision care. Sustained validation, interoperability, and physician oversight are essential for safe, effective implementation.

**Figure:**

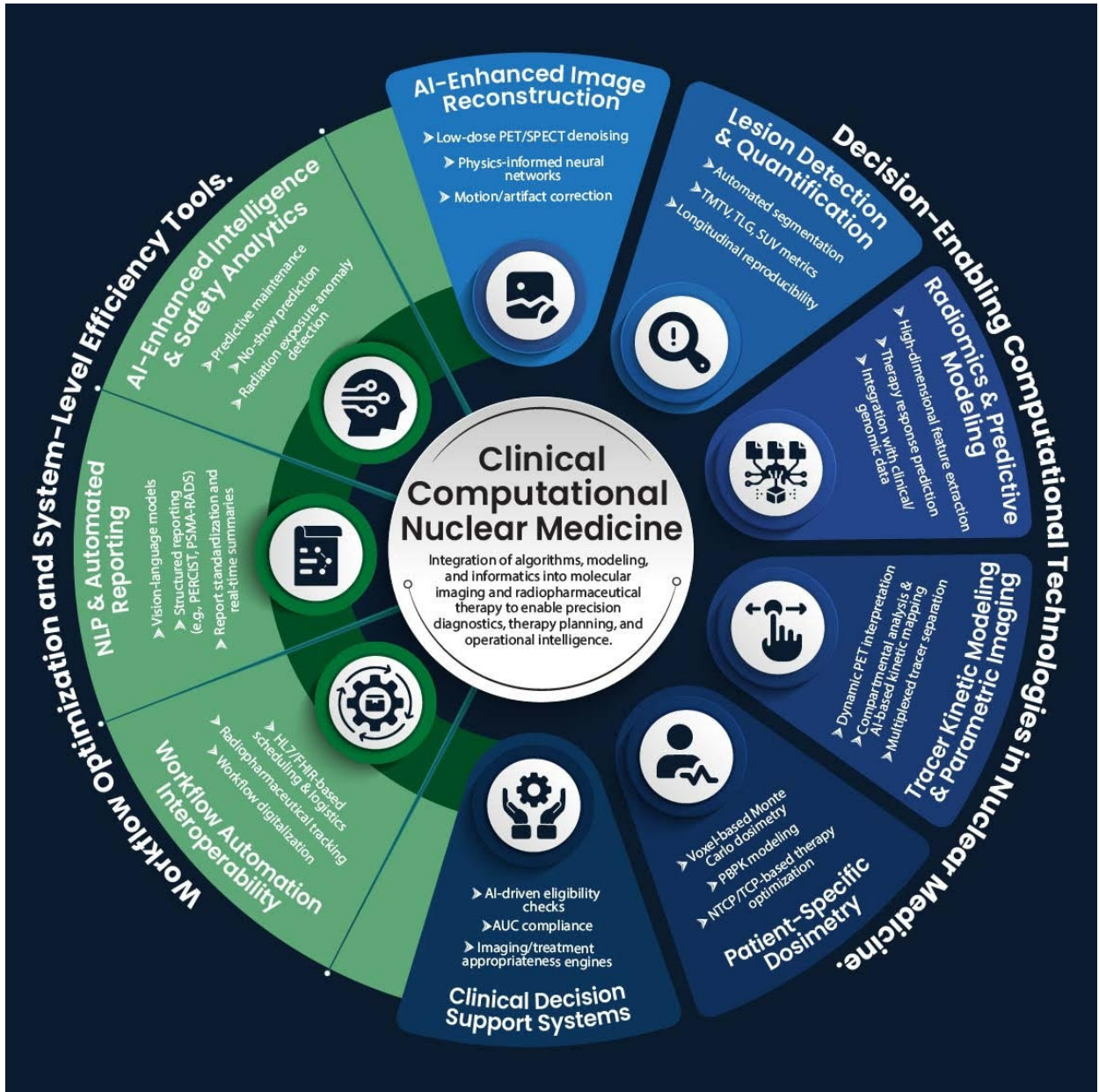

**Figure 1**. Applications of Computational Technologies in Nuclear Medicine. Key domains include decision-enabling tools and workflow optimization, highlighting the integration of AI, modeling, and informatics into nuclear medicine imaging and therapy.

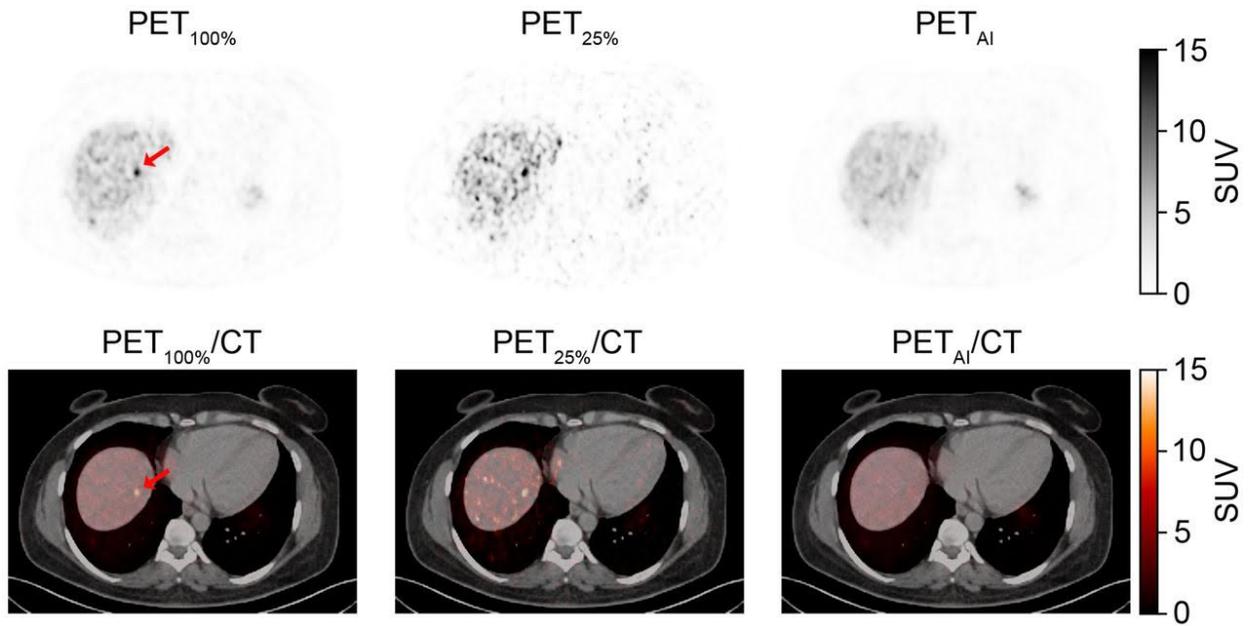

**Figure 2**. Representative case of a false-negative liver lesion on the AI-denoised PET (PET$_{AI}$) in a patient with neuroendocrine neoplasms. Although several true-positive liver lesions were concordantly identified on both the full-dose PET (PET$_{100\%}$) and PET$_{AI}$, a distinct lesion clearly visualized on PET$_{100\%}$ was not detected on PET$_{AI}$. This research was originally published in *JNM*. Loft M, *et al*. An Investigation of Lesion Detection Accuracy for Artificial Intelligence–Based Denoising of Low-Dose $^{64}$Cu-DOTATATE PET Imaging in Patients with Neuroendocrine Neoplasms. J Nucl Med. 2023;64(6):951–959. © SNMMI.[19]

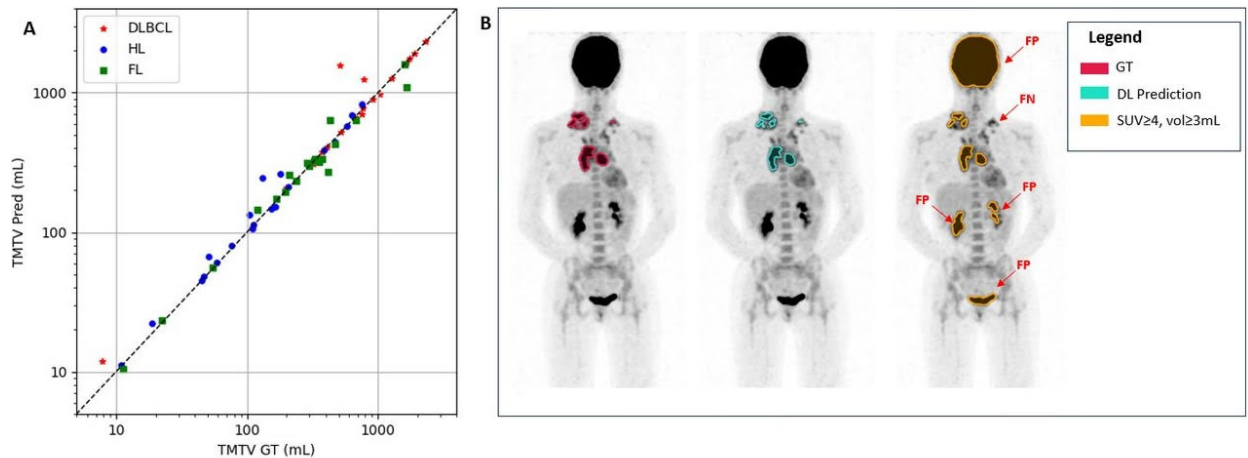

**Figure 3**. Comparison of TMTV predictions from a pretrained DL model versus ground truth in DLBCL, HL, and FL. (A) Scatterplot of predicted vs. reference TMTV values. (B) MIP images showing lesion contours: ground truth (red), DL predictions (cyan), and SUV4-based preselection (orange). DL approach demonstrated higher accuracy and fewer false positives than threshold-based methods. This research was originally published in *JNM*. Ionescu G and Willaime J. AI-assisted TMTV calculation for lymphomatous disease – validation study on the international TMTV benchmark dataset. J Nucl Med. 2025;66(Suppl 1):251795. © SNMMI.[95]

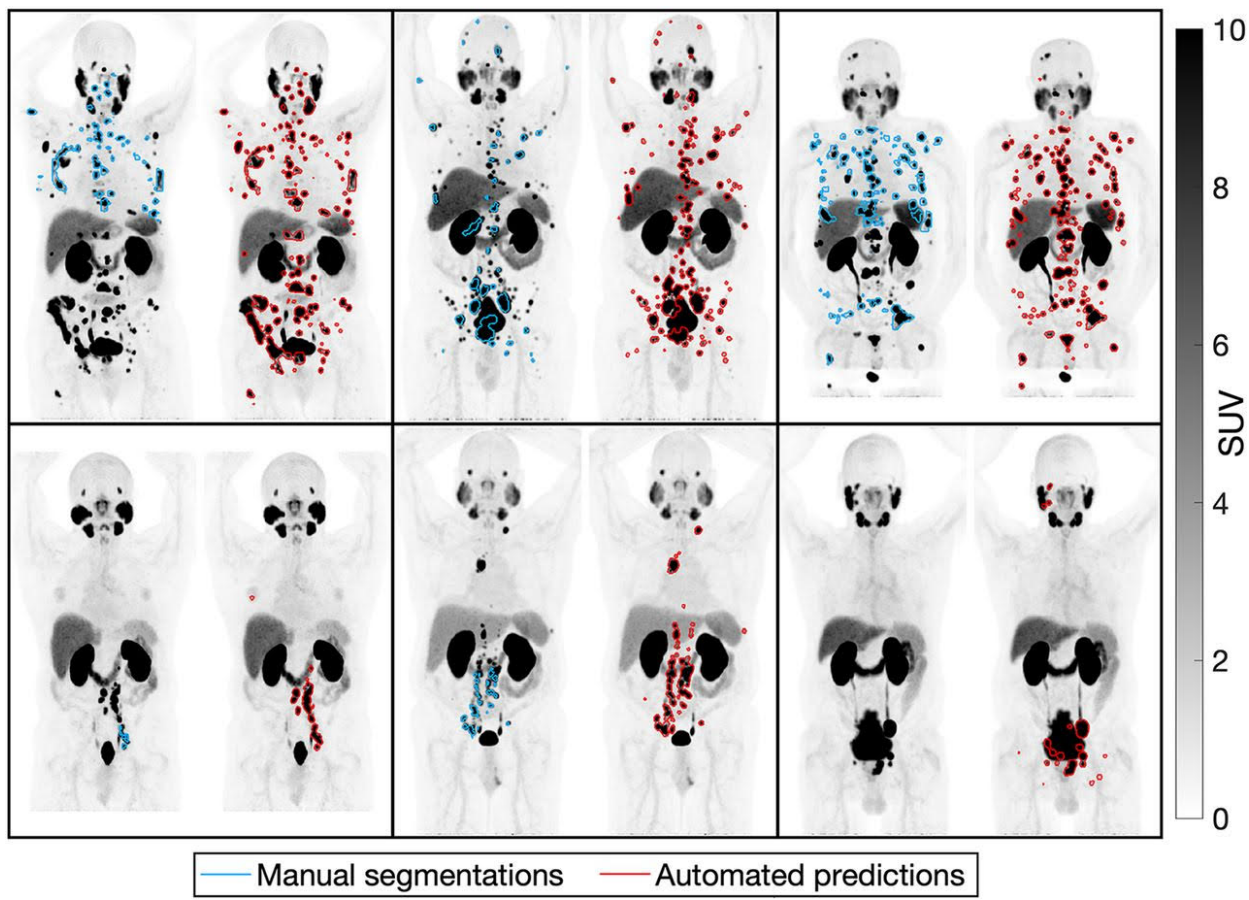

**Figure 4**. Comparison between sparse manual annotations and AI-predicted segmentations on PSMA PET scans of six patients with prostate cancer. The semisupervised DeepSSTL model, trained on incompletely labeled data, identifies additional PSMA-avid lesions not delineated in the manual reference, illustrating the model's capacity to generalize beyond partial supervision and capture whole-body tumor burden. This research was originally published in *JNM*. Leung KH. *et al.*, Deep semisupervised transfer learning for fully automated whole-body tumor quantification and prognosis of cancer on PET/CT. J Nucl Med. 2024;65(4):643–650. © SNMMI.[27]